\documentclass[runningheads]{llncs}
\usepackage{graphicx}
\usepackage{blindtext}
\usepackage{booktabs}
\usepackage{fixltx2e}
\usepackage{adjustbox}
\usepackage[T1]{fontenc}
\usepackage{makecell}

\usepackage[justification=centering]{caption}

\newcounter{example}[section]
\newenvironment{example}[1][]{\refstepcounter{example}\par\medskip
   \noindent \textbf{Example~\theexample. #1} \rmfamily}{\medskip}
   
\begin{document}
\title{Automated Approach to Improve IoT Privacy Policies}
\subtitle{Technical Report, 2018}
%
%
\author{Parvaneh Shayegh \and Vijayanta Jain \and
Amin Rabinia \and
Sepideh Ghanavati}
\authorrunning{P. Shayegh et al.}
%
\institute{University of Maine, Orono, ME, USA 
}
\maketitle              
\begin{abstract}
The massive growth of the Internet of Things (IoT) as a network of interconnected entities~\cite{RZL13}, brings up new challenges in terms of privacy and security requirements to the traditional software engineering domain~\cite{AAR2017}. To protect the individuals’ privacy, the FTC’s Fair Information Practice Principles (FIPPs)~\cite{Fed16} proposes to companies to give notice to the consumer about their data practices, provide them with choices and give them means to have control over their own data.. Using privacy policy is the most common way for this type of notices. However, privacy policies are not generally effective due to two main reasons: first, privacy policies are long and full of legal jargon which are not understandable by a normal user; second, it is not guaranteed that an IoT device behave as it is explained in its privacy policy. In this technical report, we propose and discuss our methodologies to analyze privacy policies. By the help of this analysis, we reduce the length of a privacy policy and make it organized based on privacy practices to improve understanding level for the user. We also come up with a method to find the inconsistencies between IoT devices and their privacy policies. 

\keywords{Privacy Policies \and Internet of Things (IoT) \and Supervised Machine Learning \and Topic Modeling.}
\end{abstract}
\section{Introduction}
\label{Introduction}
Data privacy has become a major challenge for companies and social network providers, particularly in the digital world where goods and services are provided to the users on the Internet. To provide the users with reliable and efficient products (whether goods or services), companies inevitably make use of the users personal data. For example, the music industry (e.g. Soundcloud, YouTube, etc.) uses the search history of its users to provide them with the type of music they are more interested in. In such circumstances, it is of great importance to notify the users about their data privacy risks and help them make rational decisions accordingly. In doing so, companies should present their data practices in a clear and transparent way. Data practices are referred to the individual's personal data that are collected, used or shared by companies. A practical way to notify the users about data practices is to provide them with the company's privacy policy prior to the selection of any goods or services. Such privacy policies allow users to choose which parts of their personal data can be shared, and which third parties can have access to their personal data. 

Despite the potential benefits of privacy policy statements prior to selection of goods and services, these privacy policies have some issues which should be addressed. First, such statements are often ineffective, long and complicated. Schaub el.~\cite{schaub2015design} discuss the reasons behind such ineffectiveness and show that privacy policies are mostly ignored by the users. Even if privacy policies are not ignored, users cannot understand them due to the complexity of privacy policy statements. To address these shortcomings, various approaches have been suggested in the literature such as \textit{platform for privacy preferences project} \cite{reagle1999platform}, \textit{text categorization} \cite{wilson2016creation} \cite{costante2012machine} \cite{ammar2012automatic}, \textit{unsupervised classification} \cite{ramanath2014unsupervised} and \textit{recognized ambiguity} \cite{reidenberg2016ambiguity}. These approaches leverage privacy analysis techniques to help users understand privacy policies more clearly. Privacy analysis techniques also provide a solid background for further process of privacy policies e.g. making them concise and comprehensive. Second, if we assume that a privacy policy is short enough to be read and it is completely understandable, the problem related to privacy policies still remains unchanged, especially for IoT devices. For IoT devices, two important challenges exist: first, it is not guaranteed that privacy policies reflect exactly practices on personal or sensitive data. These inconsistencies between privacy policies and what the code of the device appears to do is one of the challenges with respect to privacy policies. Some of the previous research address similar problem between mobile applications and their privacy policies~\cite{zimmeck2017automated} \cite{slavin2016toward} but the inconsistency between an IoT device and its privacy policy is still remained as a research gap.

The other challenge is that when IoT devices transfer personal information among each other, what does it happen to the transferred data? For example, if device A sends personal information to device B, and then the user removes personal information from device A, what happens to the personal information which is stored on device B? Thus, not only we should check the consistency between privacy policy of device A with its code, but we need to check the code of device B as well. That is, the analysis of privacy policies and mapping them to the devices' code is essential when we want to improve privacy policies. 

In this technical report, we introduce a method to analyze privacy policies and to reduce the length of a privacy policy which is the first step in identifying inconsistencies. We, also, highlight different parts of a privacy policy based on its topic and its data practices. The applicability of this solution is tested using empirical data from privacy policies of 147 IoT devices and mobile applications. In addition, we propose a process-based approach to address the problem of IoT devices inconsistencies with privacy requirements. Testing the applicability of this approach is not discussed in this technical report.

The rest of the paper is organized as follows: First, in Section~\ref{Related_Works}, we give an overview of previous work on privacy policy challenges and analysis processes. In Sections ~\ref{privacy_analysis} and ~\ref{classification}, we discuss the detailed approach to extract privacy policies and to analyze privacy polices based on classification algorithms. We, then, compare three classification methods with each other to find a solution for privacy policy analysis with higher precision in Section~\ref{classifier-evaluation}. After analyzing privacy policies, in Section~\ref{shorten}, we remove any unnecessary sentence which is not related to the user and make the privacy policy shorter and concise.
Next, we introduce a method to categorize privacy policy parts based on their topic and also we present a
visualized model for privacy policies in Section~\ref{Categorize}. 
Finally, in Section~\ref{conclusion}, we discuss the future work and conclusion of the paper.

\section{Related Works}
\label{Related_Works}
A conventional and primary approach to deal with lengthy privacy policies was the Platform for Privacy Preferences Project (P3P)~\cite{reagle1999platform}. This approach aimed to help users make rational decision about their privacy on websites without reading them. The P3P approach automatically retrieved privacy practices from websites that had standard format ~\cite{reagle1999platform}. This approach was used for a long time with the Internet Explorer (IE) browser. Although the P3P approach had many advantages (e.g. helping users understand privacy policies in an organized and simple way), it also had major problems: it limited users access to websites, it did not recognize all privacy practices and it only worked with websites which defined their privacy policies using the P3P format ~\cite{reagle2013designing}.

Wilson et al.~\cite{wilson2016creation} argue that automated annotation of privacy policies is required prior to addressing any problem related to privacy policies. They produce a dataset consisting of 115 privacy policies (267K words) with manual annotations for 23K fine-grained data practices. The dataset includes privacy policies of top five search queries in Google web browser. They use crowdsourcing technique to annotate privacy policies and to investigate whether the data is collected or shared by a third-party. They apply logistic regression, Support Vector Machine (SVM) and Hidden Markov Model (HMM) to predict the categories (i.e. labels) of privacy policy segments and compare the performance, precision, recall and F1-scores of these three classifiers. The results shows that the SVM approach has a superior performance among the others; The HMM and the logistic regression perform similarly. Although this study provides a thorough comparison of three techniques in annotating privacy policies, it does not consider ambiguities which exist in privacy policies.

Reidenberg et al.\cite{reidenberg2016ambiguity} describe that website privacy policies often contain ambiguous language that undermines the purpose and value of privacy notices for website users. They develop a theory of vague and ambiguous terms which could address privacy policies' ambiguity. They propose a method to score parts of privacy policies based on their ambiguity. The method classifies an ambiguity in "share", "collect", "retain" and "use". They apply their proposed method to privacy policies of companies providing different types of services. The comparisons indicates that food industry companies, e.g. Costco, have privacy policies that are more ambiguous than companies that are subject to some form of regulation, e.g. Bank of America. Their approach, however, only detects ambiguities but it does not take a further step in solving these ambiguities within privacy policies. The authors, however, have stated that a combination of their method with privacy policy annotation may solve this problem.

Constane et al.~\cite{costante2012machine} suggest another approach to solve the problem of reading long privacy policy statements. They use text categorization and machine learning to categorize paragraphs of privacy terms to assess their completeness with a grade. They claim that this method helps users inspect a privacy policy in a structured way and only read those paragraphs that interest them. Following their study, Zimmeck et al. introduced "privee"~\cite{zimmeck2014privee} - a method to analyze privacy policies which integrates Constane’s classification method with Wilson’s crowdsourcing method mentioned above. In doing so, the privacy policy is returned to the user if the privacy analysis results are available in the crowdsourcing repository. Otherwise, the privacy policy is automatically classified and then is returned to the user.

Ammar et al. \cite{ammar2012automatic} come up with a text categorization method to prepare a pre-processed privacy policy for further Natural Language Processing (NLP) and  Machine Learning (ML) analysis of privacy policies. They use crowdsourced annotations of around 60 privacy policies as a dataset to train the classifier. They employ logistic regression - a classic high-performance probabilistic model- to label privacy policy documents. In doing so, they classify privacy policies into two classes: absence or presence of a concept. They experiment different concepts to determine whether the company has access to personal data and users can cancel, terminate, or delete their accounts. They mention that their proposed method can be improved by increasing the number of crowdsourced privacy policies and experimenting more concepts. 

Ramanath et al.~\cite{ramanath2014unsupervised} introduce unsupervised alignment of privacy policies using Hidden Markov Models. They apply their method to a dataset consisting of crawling privacy policies of around 1000 unique websites. To label each part of the privacy policy and title them accordingly, they use Amazon crowdsourcer and train their dataset based on the crowdsourcers output. In the next step, they apply HMM to this dataset where each HMM corresponds to a topic within the privacy policy. This HMM method uses a group of keywords to distinguish the topic, such as \textit{Art}, \textit{Business} or \textit{Game}. The evaluation shows that the approach is more effective than previous classification techniques such as clustering and topic modeling. However, the shortcoming of this approach is that privacy policies are labelled manually which makes it unfeasible for larger datasets. 

Shayegh and Ghanavati \cite{espre} proposed a method to analyze privacy policy by annotating them and highlighting each term based on its role in text. They used these annotated privacy statements, to generate a graph-based view of privacy policies and show data practices in a better way to users. This graph is also useful for adding legal requirements to privacy policies. In addition, a novel way to extract notices and choices from this graph is introduced in their paper. 

Zimmeck et al. \cite{zimmeck2017automated} develop a method to check the compliance of Android applications and privacy requirements. To extract privacy requirements of each Android app, they analyze privacy policy of that app using a machine learning method. Their research has two major parts: (1) privacy policy analysis and (2) mobile application analysis. In order to prepare the dataset for this study, they crawled Google play store and downloaded around 18000 free Android apps. To analyze their privacy policies, they use a group of classification methods: to train the data, they manually annotated 120 privacy policies as a corpus; then, they picked a box of keywords as a feature for classification. They prototype various classifiers and conclude that support vector machine and logistic regression have the best performance. As for the mobile app analysis, they design an app analysis system based on Androguard -an open source static analysis tool. Using Androguard, they could investigate the calls from other services to check which part of information is shared with third-parties. At the end, they map privacy policy of each application with the output of this system to find the inconsistencies. 

Slavin et al.~\cite{slavin2016toward} propose a method that links privacy policy phrases with Application Programming Interface (API) methods that produce sensitive information. Their proposed method aims at checking the compliance between the implementation of an Android application with its privacy policy. In doing so, they crawl play store and download 50 privacy policies. Then, they extracted anthologies from these privacy policies by manually annotating them. For the purpose of mapping, they propose two violations which could be found between policies' ontology and APIs: 1) weak violations which occur when the policy describes the data practice using vague terminology, and 2) strong violations occur when the policy does not describe an app’s data collection practice. They prove that privacy policies suffer from vague phrases.

\section{Analysis of Privacy Policies}
\label{privacy_analysis}
In this section, we discuss the process of collection and analysis of privacy policies. In the next section, we describe the detailed process for analyzing, classifying and shortening privacy policies. 

\begin{figure*}[h]
\includegraphics[width=1.0\textwidth]{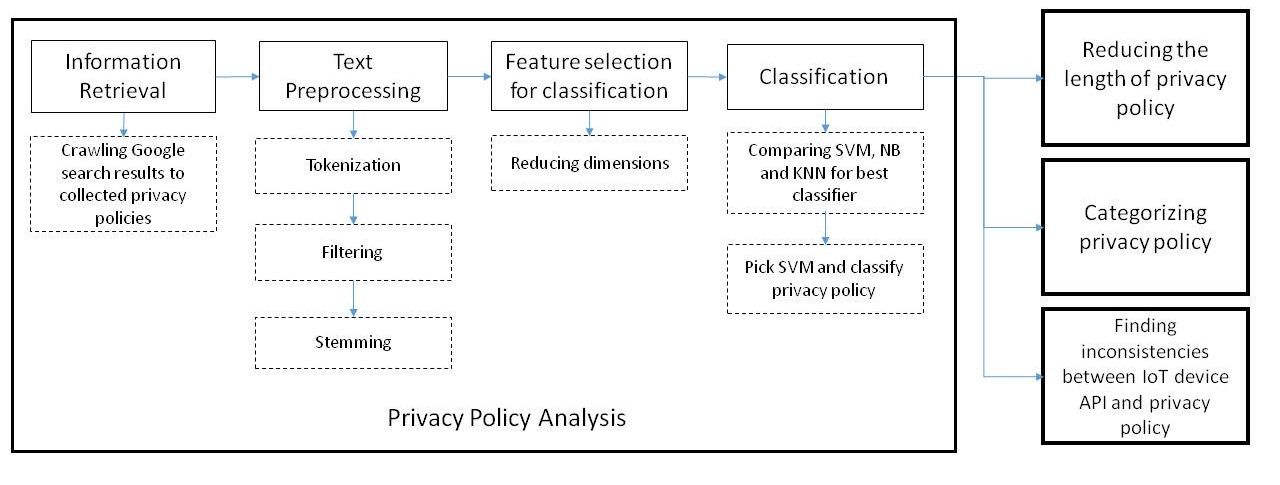}
\centering
\caption{Our Proposed Privacy Policy Analysis Approach}
\label{fig:Steps}
\end{figure*}

As discussed in Section~\ref{Introduction}, our end goal is to identify inconsistencies between privacy policies and the IoT devices and mobile applications. Before addressing this challenge, we need to extract relevant data practices from privacy policies and develop our dataset. For this, we extract, analyze and shorten privacy policies from both the IoT and mobile applications. After extracting data practices and features from privacy policies, we can use the result for the comparison step. Figure~\ref{fig:Steps} shows the steps of our approach for shortening and categorizing the privacy policies. 


\subsection{Collecting Privacy Policies}
We collected around 150 privacy policies from both IoT devices and mobile applications by crawling Google as well as the Playdrone~\cite{PlayDrone} dataset. Then, we converted these privacy policies into simple text files.

\subsection{Extracting Statements}
We define two classes of statements - \texttt{sensitive} and \texttt{non-sensitive}, that are present in privacy policies. To analyze privacy policies, the first step is to break a privacy policy into its sentences and to classify its parts by deciding whether they define any data practices or not.

We call a sentence as \texttt{sensitive} if it contains information about data practices and user's choices. In contrast, we define a sentence as \texttt{non-sensitive} if it does \textit{not} define any data practices or user's choice.   For example, \emph{"Xbox One and Kinect offer easy and approachable ways to control your games and entertainment with your voice and gestures."} \cite{kinect} is a sample sentence of Kinect privacy policy which is a definition and does not impact user's choices or does not notify about any data practice. In addition, the parts of privacy policies which describe privacy rules and laws are not related to the user and the user cannot make decision about them. Thus, we will label this statement as \texttt{non-sensitive}.

To classify each sentence in all the extracted privacy policies as \texttt{sensitive} or \texttt{non-sensitive}, we use machine-learning based classifier. To find the best classifier for this task, we test three most popular classification methods on 40 of the privacy policies that we have extracted. The selected classification methods are: K Nearest Neighbour (KNN) \cite{weinberger2006distance}, Support Vector Machine (SVM) \cite{zeng2008fast} and Naive Bayes (NB) \cite{allahyari2017text}. We, then, compare their precision with each other and choose the most precise one to use for the rest of the privacy policies.

In the next section, we discuss the classification methods for the first 40 privacy policies from our dataset.

\section{Classification}
\label{classification}
In this Section, we describe how we prepared the privacy policy statements to classify them as \texttt{sensitive} or \texttt{non-sensitive}. The first step before using classification methods is to select features and train the classifier based on them. One of the common approaches to prepare a text for classification is bag-of-words technique~\cite{guyon2003introduction}. In this technique, grammar and order of words in a text or sentence are not important. The only important factor in this model is the presence of a word and its frequency. This technique is useful in our work because we aim to use frequency of occurrence of each word as a feature for classification. One of the problems with this technique is that frequency of common words like "the", "and" and "to" is too high and this high frequency can impact the classification output. These type of words are known as stop words. In order to ensure that the these stop words do not affect the result of our classification techniques, after the information retrieval step, we pre-process our data in two steps: Tokenization and Stemming. After processing we evaluate 3 classifiers - KNN, NB, and SVM; based on the performance, we find that SVM performs better than the other two classifiers and thus we use SVM to classify policy statements. The results of classification are discussed in detail in \ref{classifier-evaluation}

subsection{Tokenization}
\label{filtering}
After extracting the sentences from the privacy policies, we tokenize the sentences into words and remove the stop words. Stop words are commonly preposition, connector, pronouns etc. In Example \ref{example1}, we remove the words \emph{we, may, with, our, other, of, and , for, the, in , this} and we keep this set of words \emph{share, personal information, family, affiliated, companies, brands, purposes, described, privacy, statement}. We follow the same process for all of the extracted privacy policies. As a result, we have a set of words $W$ which consists of these words $w_1$...$w_n$

\begin{example}
\label{example1}

``We may share Personal Information: With our family of affiliated companies and brands for the purposes described in this Privacy Statement.''
\end{example}

\begin{equation}
    W={w_1,w_2,w_3,...,w_n}    
\end{equation}

We notice that in each privacy policy around 120 sentences exist, and each sentence has 17 words on average. From these 17 words, around 2/3 are stop words which means that we have 6 keywords in each sentence and 720 keywords in a privacy policy. However, around 30\% of these words are unique which means that we have 200 words per privacy policy. On the other hand, we found out that privacy policies have many words in common since the topics of all of them are very similar as we focus on privacy policies of IoT and mobile applications which makes the variety of words shorter. Therefore, we need to process these words to get a small bag of words. 

\subsection{Stemming}
\label{stemming}
After tokenizing the sentences, we process the rest of the words by getting their root-word, a process known as stemming. Since the topics of privacy policies are not very wide, the words which are used in each privacy policy are not very distributed. As a result, a word or a different derivation of a word may be repeated several times in each privacy policy. In Example \ref{example collect}, three different sentences from Kinect \cite{kinect} are shown. In these three sentences ``collect", ``collected" and ``collects" are used to explain data practices. However, all of them are different derivation of the word ``collect".

\begin{example}
\label{example collect}
\begin{enumerate}
    \item We collect information on how your Kinect device and platform software are functioning.
    \item Voice data may be collected to enable search and to control the console.
    \item Kinect collects and uses body recognition data to enable you to control and play games.
\end{enumerate}
\end{example}

In stemming, we keep the root of the word and remove other derivations. However, for the frequency, we consider the summation of frequencies of all of the derivations. We define a function $f(w)$ which is the frequency of word $w$.

To clarify this, in Example \ref{example collect}, three forms of collect are combined together and we keep only ``collect". As a result $f(collect)$ is the summation of all three derivations of the word which means $f(collect)= 65$ for Kinect privacy policy.

Thus, we find the root of each word in W, and we remove other derivations of them. For the first 40 privacy policies, we concluded that around 700 unique words remain after stemming. Therefore, we have a vector of 700 words. This new $W$ is the set of dimensions for classification. 

\begin{equation}
    W = {w_1, w_2, w_3,..., w\textsubscript{700}}
\end{equation}

Then, we use function $f(w)$ and find the frequency of each word in $W$. Next, we store it in a vector $F$. 

\begin{equation}
    F={f(w_1),f(w_2)....f(w\textsubscript{700})}    
\end{equation}

\subsection{Feature Selection}
After processing the data, we have a bag of words model which contains all words from privacy policies, we call $W$. This model is very useful to use as features for classification. However, the huge number of words in this model (around 700 words) make classification methods unfeasible. Thus, we apply a reducing dimension technique to make features set smaller. In order to reduce dimensions, we sort the words based on the sum of occurrence frequency between all sentences, and then we keep only the first 500 words. At the end, a dataset with 500 columns (extracted words) and around 5500 rows (number of sentences from 40 privacy policies) is built. We use them as our training dataset for training the KNN, SVM and NB classification algorithms. In our comparison, we find that SVM performs better than the other two classifiers and therefore we choose SVM to classify the rest of the privacy policies based on \texttt{sensitive} and \texttt{non-sensitive} categories.

\section{Evaluation of Classifiers}
\label{classifier-evaluation}

In this section, we compare the performance of SVM, KNN and NB on our dataset in 5-fold cross-validation. Text classification has been used in different fields of study. One of these fields is to analyze privacy policies as we discussed in Section \ref{privacy_analysis}. In our approach, we aim to predict the label of each sentence in a privacy policy based on pre-defined labels which are \texttt{sensitive} data and \texttt{non-sensitive} data.

\subsubsection{NB Classifier}
The Naive Bayes classifier is one of the simplest and popular classifier which is used in many applications. It uses a probabilistic model to find the distribution of different terms and uses this distribution for classification. In text classification, NB has two different models to classify the text. We use the model that takes frequency of the words into account instead of the model which uses binary values for presence or absence of the words. 

\subsubsection{SVM Classifier}
Support Vector Machines (SVM) is widely used for text classification problems. This method is one of the linear classifier approaches which finds the best vector to separate the items of class A from class B based on their features. In this paper, we use single SVM classifier as we only have  two classes \texttt{sensitive} sentences and \texttt{non-sensitive} sentences. 

\subsubsection{KNN Classifier}
K-Nearest Neighbour classifier uses a distance measure to classify text. This method puts every two sentences in one class if they are close to each other~\cite{weinberger2006distance}. We use cosine similarity to calculate distance between each two sentences \cite{liu2004text}.

\subsubsection{Classification Step}
\label{classificationsteps}
Two metrics are widely used for evaluation of classifiers: \textit{precision} and \textit{recall}. However, in our proposed approach, we use \textit{True Negative Rate} as the third metric. In our approach, precision is the fraction of sensitive data which are labeled correctly among all sentences which are labeled sensitive with the classifier. Equation \ref{precision} shows the equation for calculating \textit{precision}. \textit{Recall} is the fraction of sensitive data which are labeled correctly among all sentences which are actually sensitive and it is shown in equation \ref{Recall}. True Negative Rate is the fraction of non-sensitive data which labeled correctly among all data which are actually non-sensitive and is shown in equation \ref{negative_rate}. 

In these equations, TP (True Positive) is the number of sentences which are sensitive and predicted as sensitive, FP (False Positive) is the number of sentences which are not sensitive but are predicted as sensitive, TN (True Negative) is the number of non-sensitive data which are predicted correctly and FN (False Negative) is the number of sensitive data which are predicted as non-sensitive. 

\begin{equation} 
\label{precision}
Precision = TP/(TP+FP)
 \end{equation}
 
 \begin{equation} 
\label{Recall}
Recall = TP/(TP+FN)
 \end{equation}
 
  \begin{equation} 
\label{negative_rate}
Negative rate = TN/(TN+FP)
 \end{equation}

As mentioned in Section~\ref{privacy_analysis}, we randomly picked 40 of the privacy policies that we extracted to analyze these three classifiers and selected the one that has better \textit{precision}, \textit{recall} or \textit{True Negative Rate}. 

In our analysis, we found that \textit{recall} for all of the cases is very close to 1. It means that by using our proposed approach for classification, non sensitive data is labeled as non-sensitive. The reason is that sensitive data have specific words in common. For example, collect, information, share, etc. However, non-sensitive data do not have specific keywords. Therefore, when classifier calculates the distance between a sentence and the data in sensitive class, it finds more similarity in compare with data in the non-sensitive class. For example, in sentence~\ref{example 2}, the true label is non-sensitive since it is a definition for Alexa Interaction and it is not related to the user. However, because of the presence of words such as information, interaction and use, classifiers label it as sensitive. 

According to this fact, we can conclude that we never lose any sensitive data while reducing the length of the privacy policy which we explain in Section \ref{shorten}. 

\begin{example}
\label{example 2}
``Alexa Interactions'' means all information related to your use of Alexa and Alexa Enabled Products.
\end{example}

\begin{figure*}[h]
\includegraphics[width=1.0\textwidth]{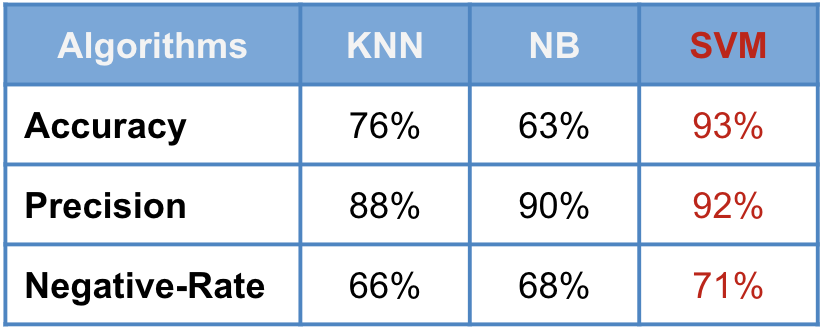}
\centering
\caption{Comparison of KNN, NB and SVM based on their Precision and Negative Rate}
\label{fig:Comparison}
\end{figure*}

After completing the classification step, we end up having some non-sensitive sentences with sensitive label which are measured by \textit{precision} and \textit{True Negative Rate}e.  Figure~\ref{fig:Comparison} presents the result for precision and true negative rate for each of the classification methods.  In fact, precision shows the percentage of sensitive sentences which are labeled correctly and true negative rate shows the percentage of non-sensitive sentences which are labeled as sensitive. 


By comparing the results, we found that SVM is a better classifier for our purpose. 

Next, we used SVM to classify the rest of the privacy policies (~107 privacy policies). The total sentences in these 107 privacy policies is 16822. Figure~\ref{fig:svm} shows the accuracy, precision, recall and F-Score for the rest of the extracted privacy policies and Figure~\ref{ResultofSensitive} shows the percentage of \texttt{sensitive} and \texttt{non-sensitive} sentences.

\begin{figure*}[h]
\includegraphics[width=1.0\textwidth]{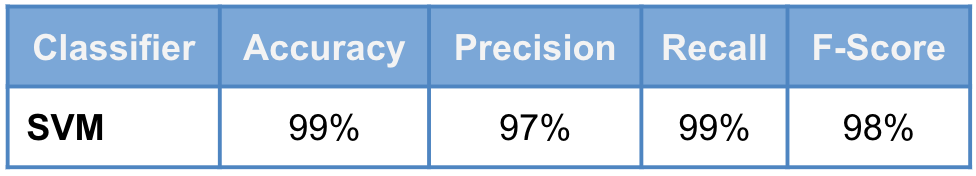}
\centering
\caption{Result from Using SVM for Classifying Privacy Policies}
\label{fig:svm}
\end{figure*} 

\begin{figure*}[h]
\includegraphics[width=1.0\textwidth]{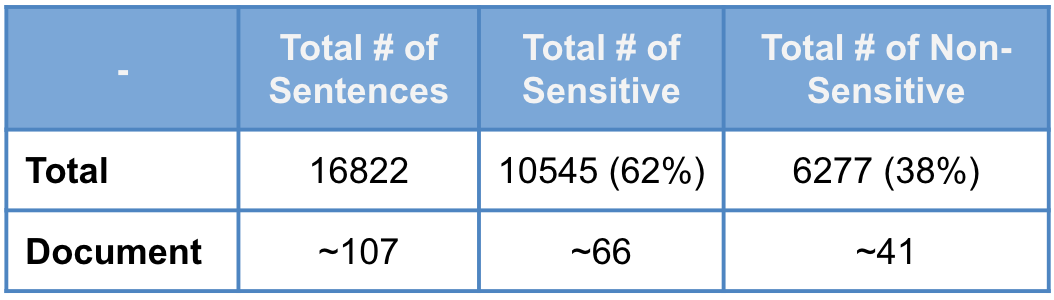}
\centering
\caption{Percentage of Sensitive vs. Non-Senesitive Sentences}
\label{fig:svm}
\end{figure*}


\section{Shorten Privacy Policies}
\label{shorten}
Reducing the length of a privacy policy, make them shorter and easier to be read and to be understood. In Section~\ref{privacy_analysis}, we divided privacy policy sentences into two groups, sensitive and non-sensitive sentences. In this section, we remove the sentences with non-sensitive label by considering the fact that these sentences are not related to the user. For example, Amazon Alexa privacy policy~\cite{amz2017} has 116 sentences in general. By removing its non sensitive sentences, the number of sentences is reduced by 38\% and the number of words are decreased from 1157 to 820 words.  

\section{Categorize Privacy Policies}
\label{Categorize}
Breaking the sensitive part of a privacy policy into sub-groups makes the privacy policy categorized and well-organized. These groups help users find relevant part of the privacy policy by the data practices and the type of information which is collected, used or shared. Furthermore, users can easily understand the purpose of the company for such data practices. For example, if a user wants to know about collecting personal data, he/she can easily refer to the \texttt{collection} category and find all the relevant sentences about personal data collection. 

To categorize privacy policies, we leverage Stanford Topic Modeling Toolbox~\cite{ramage2011stanford} which is a topic modeling method to highlight each part of the privacy policy. This toolbox uses labeled Latent Dirichlet Allocation (LDA) to discover several topics in a document. To use this toolbox, we use the annotated, tokenized and stemmed text from Sections~\ref{classification} and ~\ref{classifier-evaluation} as an input for the topic modeling. The topic modeling method performs better on meaningful words and thus irrelevant words must be removed. To address this problem, \cite{ramage2011stanford} suggests to remove those words that appear in less than four privacy policies. 

We need to define the topics prior to the use of LDA topic modeling in this toolbox. The aim is to find the topics of sentences within the sensitive class based on the required data practice type. We define these sub-classes as: \texttt{Information}, \texttt{Collection}, \texttt{Sharing}, \texttt{Permission} and \texttt{Technology}. 

\begin{table}[]
\centering
\caption{Definitions of Topics in Privacy Policies}
\label{definition}
\begin{tabular}{|p{2cm}|p{6.5cm}|p{3.5cm}|}
\hline
\thead{Topics} & \thead{Definition} & \thead{Examples}  \\ \hline
Information & Any sentence which contains personal information. & Personal data, Email, Audio, Mailing Address \\ \hline
Collection & Any sentence which is related to collection of data. & Collect, Access, Use, Store \\ \hline
Sharing & Any sentence which explains sharing data with other parties. & Disclose, Share, Reveal, 3rd Party \\ \hline
Permission & The parts of the privacy policy that needs permission of the user to collect, share or use user's data. & Agree, Consent, Allow, Permit\\ \hline
Purpose & The sentences about the purpose of the data practices. & Purpose, To Provide, To Help, To Offer\\ \hline
Technology & Any sentence which explains technological aspects of data collection or sharing. & Cookie, Device, Session, Service\\ \hline
\end{tabular}
\end{table}

The above mentioned topics are shown in Table~\ref{definition}. In each row, the first word is the topic and the rest of the words are a small sample of meaningful words that we use to distinguish the corresponding topic. Table~\ref{ex-topic} shows three random sentences with their topics. As it is shown in this table, a sentence must be assigned to at least one topic but it may be assigned to multiple topics as well. 

\begin{figure*}[h]
\centering
\includegraphics[width=1.0\textwidth]{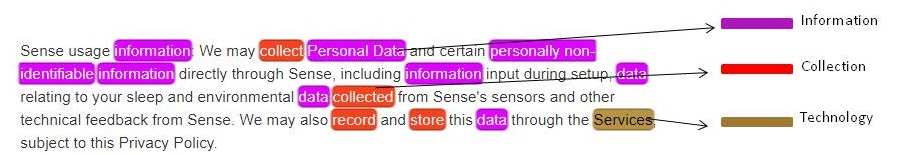}
\caption{An Example of a Text with Each Topic}
\label{textTopic}
\end{figure*}

\begin{figure}[h]
\centering
\includegraphics[width=8cm]{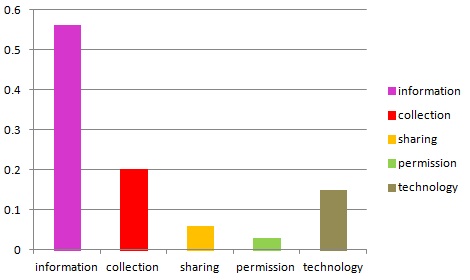}
\caption{The Distribution of Topics in our Dataset}
\label{chart}
\end{figure}

In Figure \ref{textTopic}, we selected a paragraph from Hello Inc. privacy policy \cite{Hello} and highlighted three topics based on their corresponding words to clarify how the proposed method recognizes the topics in a text. These three topics are \texttt{Information}, \texttt{Collection} and \texttt{Technology}. The distribution of these topics in the data is shown in Figure~\ref{chart}. The results indicate that Information is a topic which is used in more than half of privacy policies whereas Permission is a topic which is ignored by most of the privacy policies. 

\begin{figure}[h]
\centering
\includegraphics[width=8cm]{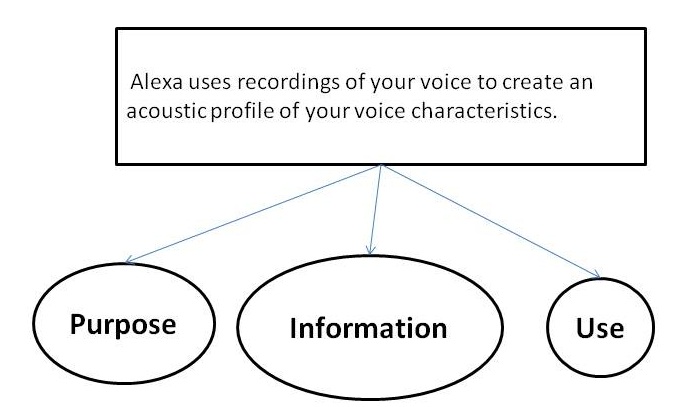}
\caption{Proposed Graph to Visualize Data Practices}
\label{graph}
\end{figure}

Also, we can use the categorized privacy policies to visualize data practices. Shayegh and Ghanavati~\cite{espre} propose a method to convert an annotated privacy policy to a graph. Their proposed graph shows links between sensitive terms. We leverage their approach and combine it with our categorized text. In Figure~\ref{graph}, a  sample  sentence  from  Amazon  Alexa  privacy  policy  is picked. As it is shown in the graph, this sentence is related to three categories use, information and purpose.

\begin{table*}[]
\centering
\caption{Example of Three Sentences from LG Smart Home Devices'~\cite{lg} Privacy Policy with Their Topics}
\label{ex-topic}
\begin{tabular}{|p{9cm}|p{3cm}|}
\hline
\multicolumn{1}{|c|}{\textbf{Sentence}}                                                                                      & \textbf{Topic(s)}     \\ \hline
This information may be used to deliver Products or service which you have purchased                                         & Information / Collect \\ \hline
We may collect your first and last name, mailing address, and usage data in order to track your usage of Products or Service & Collect/ Information  \\ \hline
We generally do not share with third parties the information we receive as a result of you using the LG Smart Home Service.  & share/ information    \\ \hline
\end{tabular}
\end{table*}

\section{Conclusion and Future Work}
\label{conclusion}
Privacy is one of the main concerns with the growth of mobile applications and IoT devices. IoT devices' users are more concern about their privacy since IoT devices can access massive amount of personal and sensitive information about the user. On the other hand, manufacturers of IoT devices are obliged to notify users about their privacy practices. The most common way for this notification is privacy policy, however, privacy policies are long, complicated and most of them do not behave exactly the same as the device does.

In this technical, we proposed a method to analyze privacy policies based on classification methods and then used the result of this analysis to make privacy policies more concise. Also, we propose a method for making privacy policies structured and categorized to help user understand privacy policies better. As a future work, we will implement this proposal. 

In future, we will also develop a tool to analyze IoT device's code. By mapping the processed code to analyzed privacy policy, we aim to identify inconsistencies and propose resolution strategies for them. Identifying and resolving these inconsistencies assist developers to ensure their code to be compliant with privacy policy. In addition, having privacy policies that are consistent with the IoT device data practices, can help building trust between the users and the companies.

Our proposed approach has some drawbacks which should be addressed in future work. First, in feature selection approach, we ignore the order of the words and remove all the stop words without considering their roles in the sentence and thus, it can result in some incorrect results. For example, in the sentence "we don't collect information", it is important to consider the word "do not" before the verb. One of the solution we aim to use in our future work is using N-gram model to store this spatial information within the text. We also keep only the most frequent words as dimensions for classification but we may lose some important none-frequent words in this way. One solution is to keep these words instead of removing them and use a classifier with higher performance. 

Second, the number of IoT and mobile privacy policies that we have used for testing our approach is still not enough. In future, we aim to automate the extraction of privacy policies and build a larger dataset to ensure our results are more precise. 

%
%
%
\bibliographystyle{splncs04}
\bibliography{references}
%




\end{document}